\documentclass[journal,12pt,onecolumn,draftclsnofoot,]{IEEEtran}
\usepackage{amsmath,latexsym,amssymb,stmaryrd}
\usepackage{setspace,stackrel,tikz,graphicx}
\usepackage{exscale,relsize,subfig,textcomp,stackrel,setspace,float}
\usepackage{mdframed,pifont}
\usepackage{multicol,xfrac}
\usepackage{multirow}
\usepackage{mathtools}
\usepackage[T1]{fontenc}
\usepackage{slashbox}
\usepackage[font=footnotesize]{caption}
\usepackage{lipsum}
\usepackage{multicol}
\usepackage{graphicx}

\begin{document}

\title{\vspace{0cm} \LARGE A Framework for Amalgamating Optical-OFDM and Optical-OFDM with Index Modulation for Optical Wireless Systems\vspace{0em}}
\makeatletter
\makeatother
\author{Ali Waqar Azim,
        Yannis Le Guennec, Laurent Ros,
        and~Marwa Chafii
\thanks{Ali Waqar Azim is with Department of Telecommunication Engineering, University of Engineering and Technology, Taxila, Pakistan (email: aliwaqarazim@gmail.com)}
\thanks{ Yannis Le Guennec and Laurent Ros are with Universit\'{e} Grenoble Alpes, CNRS, Institute of Engineering, Grenoble INP, GIPSA-LAB, \(38000\) Grenoble, France (email: \{yannis.le-guennec,laurent.ros\}@grenoble-inp.fr).}
\thanks{Marwa Chafii is with Engineering Division, New York University (NYU) Abu Dhabi, 129188, UAE and NYU WIRELESS, NYU Tandon School of Engineering, Brooklyn, 11201, NY, USA (email: marwa.chafii@nyu.edu)}}

\maketitle
\begin{abstract}
In this communication, we propose a framework for amalgamating optical-orthogonal frequency-division multiplexing (O-OFDM) and O-OFDM with index modulation (O-OFDM-IM) for optical wireless systems. Both schemes individually have some limitations, e.g., O-OFDM does not provide any granularity for spectral efficiency (SE)/energy efficiency (EE) trade-off, and O-OFDM-IM loses EE for higher order modulation alphabets. By combining O-OFDM and O-OFDM-IM together, the above limitations can be circumvented. Following on from providing the general framework, we use asymmetrically clipped (AC)O-OFDM and ACO-OFDM-IM to demonstrate the validity of the framework. The results presented herein establish that, by consolidating O-OFDM and O-OFDM-IM, in addition to providing SE/EE trade-off, we may realise higher SE and outperform O-OFDM-IM in terms of EE for high SE.
\end{abstract}
\IEEEpeerreviewmaketitle
\vspace{-2mm}
\section{Introduction}
\IEEEPARstart{O}{ptical}-orthogonal frequency division multiplexing (O-OFDM) based on intensity-modulation and direct-detection (ItM/DD) have been widely studied for optical wireless systems (OWS). The classical O-OFDM approaches are direct current (DC) O-OFDM and asymmetrically clipped (AC) O-OFDM \cite{carruthers1997modeling,armstrong2006power}. These techniques differ in the manner a non-negative intensity signal is obtained for transmission via ItM/DD. To be more precise, DCO-OFDM uses a bias, whereas the anti-symmetric signal in ACO-OFDM allows zero-level clipping to eliminate the negative amplitude excursions. However, O-OFDM schemes do not allow any granularity to target different spectral and energy efficiencies. For example, considering \(N\) sub-carriers and modulation alphabet cardinality of \(M\), the spectral efficiency offered by DCO-OFDM and ACO-OFDM is \((\sfrac{N}{2}-1)\log_2(M)\) and \((\sfrac{N}{4})\log_2(M)\), respectively. This is a substantial limitation of O-OFDM as it offers spectral efficiency (SE)/energy efficiency (EE) trade-off but with no granularity, which is vital for internet-of-thing applications.

One method adopted in OWS from the radio-frequency literature to improve the SE and to impart granularity for SE/EE trade-off is to use index modulated O-OFDM (O-OFDM-IM). Several O-OFDM-IM approaches have been proposed in the literature, e.g., \cite{bacsar2015optical,azim2020cl,azim2020access,azim2021filtered,azim2020laco}. In \cite{bacsar2015optical}, the authors amalgamate IM with DCO-OFDM and ACO-OFDM, resulting in approaches referred to as DCO-OFDM-IM and ACO-OFDM-IM, respectively. Both approaches can provide SE/EE trade-off, however, the increase in SE relative to the non-IM counterparts is minimal. To overcome this limitation, in \cite{azim2020cl,azim2020access}, we proposed enhanced IM variants of DCO-OFDM and ACO-OFDM, which substantially increase the SE relative to non-index modulated counterparts. Furthermore, in \cite{azim2021filtered} and \cite{azim2020laco}, we proposed filtered ACO-OFDM-IM and Layered ACO-OFDM-IM to improve the SE over the other IM and non-IM counterparts. Nonetheless, O-OFDM-IM is energy-efficient compared to O-OFDM only for low alphabet cardinalities \cite{azim2020access}. This is a notable limitation because it confines the use of these approaches merely for low spectral efficiencies.

From the aforementioned discussion, it is clear that using either O-OFDM or O-OFDM-IM independently may not be beneficial, therefore, in this communication, we provide a framework for amalgamating the above approaches. Consolidation of O-OFDM and O-OFDM-IM at the transmitter is quite simple, however, the issue is at the receiver side where the differentiation between the O-OFDM and O-OFDM-IM symbols is necessary. To overcome this issue, we provide a simple yet effective approach which uses look-up-tables (LUTs). To the best of our knowledge, the literature lacks such a framework, and this is the first ever study which provides an effective way to do so. As a proof-of-concept, we merge ACO-OFDM and ACO-OFDM-IM symbols. One limitation with ACO-OFDM and ACO-OFDM-IM is that the clipping operation generates out-of-band harmonics, which culminate in spectral overflow and increase the base-band bandwidth requirement. The increase in baseband bandwidth and enlargement of sub-carrier spacing because of odd sub-carrier activation leads to low achievable spectral efficiencies. Therefore, we adopt the filtering method proposed in \cite{mazahir2019achievable} to eliminate the out-of-band harmonics. Simulation results provided in the communication reveal the following concrete advantages of consolidating both approaches: (i) higher spectral efficiencies relative to IM and non-IM counterparts are realised using lower order alphabet cardinalities; (ii) granularity for SE/EE trade-off is obtained; and (iii) higher energy-efficiency is attained relative to other O-OFDM-IM counterparts for high SE range (greater than 1 bits/s/Hz). 

We have organised the rest of the article as follows. Section II explains a comprehensive framework for consolidating and differentiating the O-OFDM and O-OFDM-IM symbols at the transmitter and receiver, respectively. Section III provides the proof-of-concept using ACO-OFDM and ACO-OFDM-IM, and Section IV analyses the SE, SE/EE trade-off, and bit error rate (BER) performance and compares them with the classical counterparts. We render conclusions in Section V.
\section{Framework for consolidating O-OFDM and O-OFDM-IM}
\begin{figure}[tb]\centering
\includegraphics[trim={0.2 4.5 0 0},clip,scale=1.2]{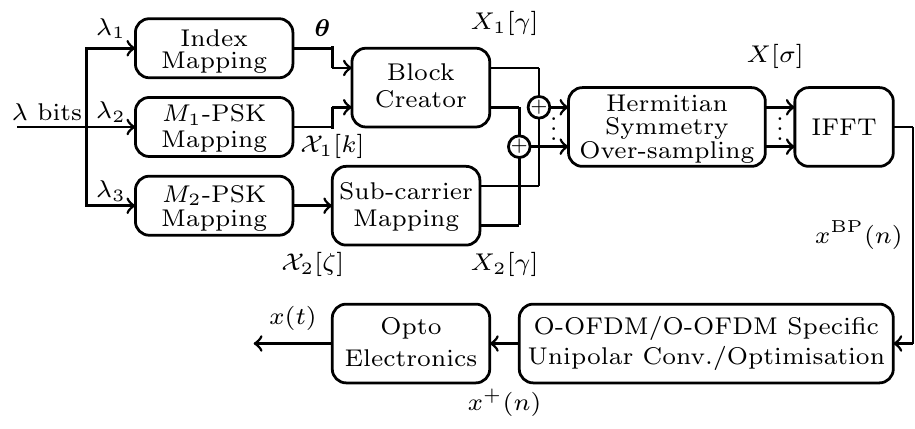}
  \caption{Transmitter architecture for the proposed framework.}
\label{tx_configuration}
\end{figure}
The transmitter configuration to be used for consolidating O-OFDM and O-OFDM-IM is presented in Fig. \ref{tx_configuration}. \(N\) sub-carriers are considered for O-OFDM and O-OFDM-IM. Moreover, \(\Omega\) sub-carriers are available for IM in O-OFDM-IM among which \(\kappa\) are activated and carry \(M_1\)-PSK alphabets, whereas, in O-OFDM, \(\Omega\) sub-carriers are modulated using \(M_2\)-PSK. Then, the total number of bits transmitted per symbol of duration \(T_\mathrm{s}\) is
\begin{equation}\label{eq01}
\lambda =\underbrace{\left \lfloor \log_2\binom{\Omega}{\kappa}\right \rfloor}_{:=\lambda_1} + \underbrace{\log_2(M_1^\kappa)}_{:=\lambda_2} + \underbrace{\log_2(M_2^{\Omega})}_{:=\lambda_3},
\end{equation}
where \(\lfloor \cdot \rfloor\) is the floor function and \(\binom{\cdot}{\cdot}\) is the binomial coefficient. It may be noticed that \(\Omega\) is equal to \(\sfrac{N}{2}-1\) and \(\sfrac{N}{4}\) for DCO-OFDM/DCO-OFDM-IM and ACO-OFDM/ACO-OFDM-IM, respectively. It is highlighted that \(\lambda_3\) bits are transmitted via O-OFDM and the increase in the number of transmitted bits relative to O-OFDM is because of \(\lambda_1 + \lambda_2\) additional bits. 

The log-concave nature of the binomial coefficient allows to approximate \(\kappa\), i.e., \(\kappa_\mathrm{approx}\) which maximises \(\lambda\) per transmitted symbol. Following \cite{azim2020access}, \(\kappa_\mathrm{approx}\) for \(M_1\) is given as:
\begin{equation}\label{eq2}
\kappa_{\mathrm{approx}}\approx \left \lfloor \frac{M_1\Omega}{M_1 + 1} \right \rfloor.
\end{equation}
Owing to (\ref{eq2}), we can attain the maximum number of bits per transmitted symbol when \(\kappa_{\mathrm{approx}}\) sub-carriers are activated for O-OFDM-IM.

\begin{figure}[tb]\centering
\includegraphics[trim={0 0 0 0},clip,scale=1.2]{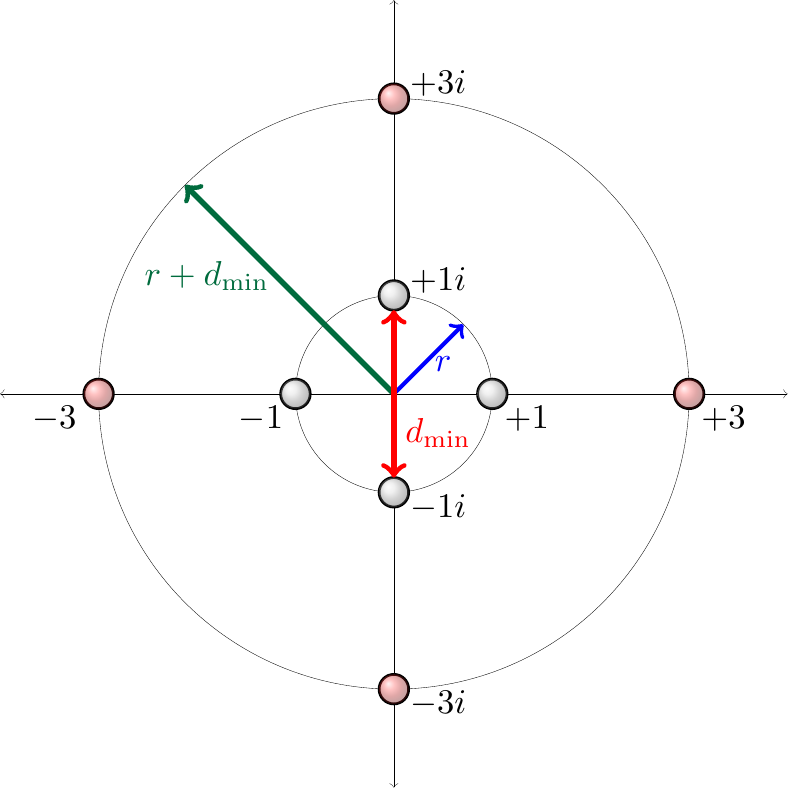}
  \caption{An illustration of constellation points used for O-OFDM (gray) and O-OFDM-IM (pink) considering \(M_1=M_2=4\).}
\label{const_fig}
\end{figure}

As mentioned earlier, the equiprobable bit sequence \(\lambda\) is parsed into three bit sequences \(\lambda_1\), \(\lambda_2\) and \(\lambda_3\). \(\lambda_1\) and \(\lambda_2\) determine the sub-carrier activation pattern (SAP), \(\boldsymbol{\theta}=\{\theta_1, \theta_2,\cdots, \theta_\kappa\}\)  with \(\theta_k \in \mathbb{Z} = \{1,2,\cdots, \Omega\}\), and \(\kappa\) \(M_1\)-PSK alphabets, \(\mathcal{X}_1[k] \in \boldsymbol{\mathcal{P}}_1\), respectively to be modulated onto the sub-carriers activated in O-OFDM-IM, where \(k \in \mathbb{K} = \{1,2,\cdots, \kappa\}\), and \(\boldsymbol{\mathcal{P}}_1\) is the set of modulation alphabets that can be used for O-OFDM-IM. \(\lambda_3\) is used to generate \(\Omega\) \(M_2\)-PSK alphabets, \(\mathcal{X}_2[\zeta]\in \boldsymbol{\mathcal{P}}_2\) for O-OFDM, where \(\zeta \in \mathbb{Z} =\{1,2,\cdots,\Omega \}\) and \(\boldsymbol{\mathcal{P}}_2\) is the set of modulation alphabets that could be employed for O-OFDM. To differentiate between the O-OFDM and O-OFDM-IM symbols in the receiver, it is imperative that a scaled PSK constellation is used for either O-OFDM or O-OFDM-IM. Hereby, we use a scaled constellation for O-OFDM-IM. Without loss of generality, considering the simplest case where \(M_1=M_2=4\), i.e., QPSK, constellation points in \(\boldsymbol{\mathcal{P}}_2\) lie on the unit circle whose radius is \(r=1\) and the minimum  distance between any pair of constellation points on the Cartesian plane is \(d_\mathrm{min}=2\). The scaling for the QPSK constellation set \(\boldsymbol{\mathcal{P}}_1\) should be such that the constellation points lie on the circle, whose radius is \(r+d_\mathrm{min} = 3\). So, the QPSK alphabets for O-OFDM-IM are chosen from the set \(\boldsymbol{\mathcal{P}}_1=\{+3,+3j,-3j,-3\}\), whilst the QPSK alphabets for O-OFDM are chosen from the set \(\boldsymbol{\mathcal{P}}_2=\{+1,+1j,-1j,-1\}\), where \(j^2 \triangleq -1\). An illustration of \(\boldsymbol{\mathcal{P}}_1\) and \(\boldsymbol{\mathcal{P}}_2\) used for O-OFDM-IM and O-OFDM, respectively is provided in Fig. \ref{const_fig}.

For \(\zeta \in \mathbb{Z}\), the \(\kappa\) \(M_1\)-PSK symbols, \(\mathcal{X}_1[k]\) (i.e., \(\mathcal{X}_1[1],\mathcal{X}_1[2],\cdots, \mathcal{X}_1[\kappa]\)) are modulated onto sub-carriers among \(\sfrac{N}{2}-1\) possible indexed by \(\zeta \in \mathbb{Z} \) as: 
\begin{equation}\label{eq}
\dot{\mathcal{X}}_1[\zeta]= \begin{cases}
			\mathcal{X}_1[k], &\zeta = \theta_k \in \boldsymbol{\theta} \\
0, &  \text{otherwise} (\text{if } \zeta \in \mathbb{Z}-\boldsymbol{\theta})
		\end{cases},
\end{equation}

To adhere with the frequency-domain frame structure of the given O-OFDM scheme, \(\dot{\mathcal{X}}_1[\zeta]\) and \(\mathcal{X}_2[\zeta]\) are modulated onto the sub-carriers as:
\begin{equation}\label{eq3}
X_1[\gamma]= \begin{cases}
			\dot{\mathcal{X}}_1[\zeta], &\zeta \in \mathbb{Z} \\
0, &  \text{otherwise}
		\end{cases},
\end{equation}
and
\begin{equation}\label{eq4}
X_2[\gamma]= \begin{cases}
			\mathcal{X}_2[\zeta], &\zeta \in \mathbb{Z} \\
0, &  \text{otherwise}
		\end{cases},
\end{equation}
where \(\gamma = \zeta \in \{1,2,\cdots, \sfrac{N}{2}-1\}\) for DCO-OFDM/DCO-OFDM-IM and \(\gamma = 2\zeta -1 \in \{1,3,\cdots,\sfrac{N}{2}-1\}\) for ACO-OFDM/ACO-OFDM-IM.

For \(\gamma \in \Gamma = \{1,2,\cdots, \sfrac{N}{2}-1\} \), the frequency-domain symbol after consolidation of \(X_1[\gamma]\) and \(X_2[\gamma]\) is expressed as:
\begin{equation}\label{eq5}
X_{1,2}[\gamma]= \sum_{u=1}^{2}X_u[\gamma].
\end{equation}

\(X_{1,2}[\gamma]\) undergoes frequency zero-padding to produce time-domain interpolation by a factor \(L\) resulting in \(X[\sigma]\) with \(\sigma \in \boldsymbol{\sigma}= \{0,\underbrace{1,\cdots, \sfrac{N}{2}-1}_{\boldsymbol{\sigma}^+},\sfrac{N}{2},\cdots,\underbrace{LN-\sfrac{LN}{2}+1,\cdots,\sfrac{LN}{2}-1}_{\boldsymbol{\sigma}^-}\}\), where \(\boldsymbol{\sigma}^+=\Gamma\) corresponds to \(\sfrac{N}{2}-1\) values for \(>0\) physical frequencies, i.e., \(\sigma\Delta f\), and \(\boldsymbol{\sigma}^-\) refers to \(\sfrac{N}{2}-1\) values for \(<0\) physical frequencies, i.e., \((\sigma-LN)\Delta f\) to be used for Hermitian symmetry incorporation with \(\Delta f = \sfrac{1}{T_\mathrm{s}}\). Explicitly, \(X[\sigma]\) is given as:
\begin{equation}\label{eq6}
X[\sigma]= \begin{cases}
			X_{1,2}[\sigma], &\sigma \in \boldsymbol{\sigma}^+\\
0, &  \text{otherwise}\\
X_{1,2}^\ast[-\sigma+LN], &\sigma \in \boldsymbol{\sigma}^-
		\end{cases},
\end{equation}
where \((\cdot)^\ast\) denotes the complex conjugate operator. Subsequently, \( X[\sigma]\) is processed using \(LN\)-order inverse fast Fourier transform (IFFT) to attain the bipolar time-domain signal, \(x^{\mathrm{BP}}(n)\) for \(n\in\{1,2,\cdots,LN\}\). \( X[\sigma]\) is scaled such that the average symbol energy of \(x^{\mathrm{BP}}(n)\) is equal to unity, i.e., \(\sum_{n}\vert x^\mathrm{BP}(n) \vert^2 = 1\). Subsequently, after scheme specific unipolar conversion, a non-negative time-domain signal, \(x^+(n)\) is attained. The average electrical symbol energy of \(x^+(n)\) is \(E_{\mathrm{s_{(elec)}}}=\sum_{n}\vert x^{+}(n)\vert^2 \). After digital-to-analogue conversion of \(x^+(n)\) and direct modulation of the bias current of an optical source (laser or light emitting diode) the intensity waveform, \(x(t)\geq 0\) is obtained and is transmitted to the optical wireless channel.

As mentioned earlier, the unipolar conversion process is specific to each modulation approach. DCO-OFDM/DCO-OFDM-IM requires addition of a bias and subsequent clipping in case there are still some negative amplitude excursions after bias addition. On the other hand, because of frequency-domain frame structure of \(X[\sigma]\) in ACO-OFDM/ACO-OFDM-IM, \(x^\mathrm{BP}(n)\) obeys the anti-symmetric property, i.e., \(x^\mathrm{BP}(l)=-x^\mathrm{BP}(l+\sfrac{LN}{2})\) for \(l=\{1,2,\cdots,\sfrac{LN}{2}\}\), therefore, the unipolar conversion only requires clipping operation.
\begin{figure}[tb]\centering
\includegraphics[trim={0.2 0 0 0},clip,scale=1.2]{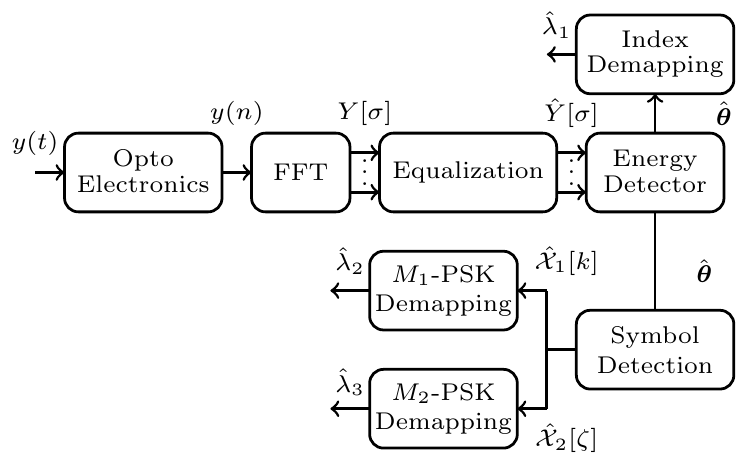}
 \caption{Receiver architecture for the proposed framework.}
\label{rx_configuration}
\end{figure}

We present the receiver architecture for the proposed framework in Fig. \ref{rx_configuration}. After analogue-to-digital conversion of the received waveform, \(y(t)\), the time-series signal, \(y(n)\) is attained. Then using \(LN\)-order FFT and zero-forcing equalisation, we attain \(\hat{Y}[\sigma]\), from which the useful information is extracted as \(\dot{Y}[\zeta]=\alpha\hat{Y}[\gamma]\), where \(\alpha\) is equal to \(1\) and \(2\) for DCO-OFDM/DCO-OFDM-IM and ACO-OFDM/ACO-OFDM-IM, respectively. Ideally, the sub-carriers carrying both O-OFDM and O-OFDM-IM alphabets have higher energies compared to the sub-carriers having only the O-OFDM alphabets. Therefore, a forthright approach to determine the SAP is to evaluate the energies as \(\xi_\zeta=\left \vert \hat{\dot{Y}}[\zeta]\right\vert^2\) and to select the indices of the sub-carriers having \(\kappa\) highest energies: which when sorted in descending order gives \(\hat{\boldsymbol{\theta}}\), from which, the bit sequence \(\hat{\lambda}_1\) is determined. Note that the sub-carriers might have either O-OFDM symbols or both O-OFDM and O-OFDM-IM symbols. By determining \(\hat{\boldsymbol{\theta}}\), we have an estimate which sub-carriers have both O-OFDM and O-OFDM-IM symbols. So, firstly only these sub-carriers are processed, using which we identify the entry of LUT which has the least Euclidean distance to the received superimposed symbols. Note that for \(M_1\)-PSK and \(M_2\)-PSK, a LUT having \(M_1M_2\) entries is needed. Thereafter, the sub-carriers having only O-OFDM symbols are processed. Once all the sub-carriers are processed, an estimate of the O-OFDM-IM and O-OFDM alphabets, i.e., \(\hat{\mathcal{X}}_1[k]\) and \(\hat{\mathcal{X}}_2[\zeta]\) is attained, from which, the bit sequences \(\hat{\lambda}_2\) and \(\hat{\lambda}_3\) are determined.

Considering \(M_1 = M_2 = 4\), i.e., QPSK for both O-OFDM and O-OFDM-IM, an example of LUT is provided in Table 1. It may also be noticed that \(\kappa\leq \Omega\), therefore, only \(16\) possibilities exist (using QPSK)  originating from both O-OFDM and O-OFDM-IM symbols on the sub-carriers (cf. Table 1). Thus, these possibilities are interpolated into a LUT, which are considered to be known at the receiver. 
\begin{table}[tb]
\caption{An example of LUT for \(M_1=M_2=4\). }
{\begin{tabular}{|l|l|l|l|l|}\hline
\backslashbox{{\(X_2[\gamma]\)}\kern-4em}{\kern-1em {\(X_1[\gamma]\)}} & \(+3\) & \(+3j\)& \(-3j\)& \(-3\)\\\hline
\(+1\) & \(+4\) & \(+1+j\) & \(+1-3j\) & \(-2\)\\\hline
\(+1j\) & \(+3+1j\) & \(+4j\) & \(-2j\) & \(-3+1j\)\\\hline
\(-1j\) & \(+3-1j\) & \(+2j\) & \(-4j\) & \(-3-1j\)\\\hline
\(-1\) & \(+2\) & \(-1+3j\) & \(-1-3j\) & \(-4\)\\\hline
\end{tabular}}{}
\end{table}
\section{Proof-of-Concept for the Proposed Framework} To validate the proof-of-concept of the proposed framework, we use ACO-OFDM and ACO-OFDM-IM. We refer to the approach resulting from the amalgamation of these schemes as Hybrid O-OFDM-IM (HO-OFDM-IM) in the sequel. It is recalled that only the odd sub-carriers can be modulated in ACO-OFDM/ACO-OFDM-IM to attain an anti-symmetric time-domain signal which is clipped to attain a non-negative signal. However, clipping of \(x^\mathrm{BP}(n)\) results in spectral overflow, thus, the baseband bandwidth of the clipped signal, \(x^\mathrm{c}(n)\) is increased to \(\tilde{B}\approx \sfrac{N}{T_\mathrm{s}}\) \cite{mazahir2019achievable} relative to \(B_\mathrm{opt}=\sfrac{N}{2T_\mathrm{s}}\): which is the baseband bandwidth of a real-valued bipolar pure frequency tones based signal, where \(\Delta f = \sfrac{1}{T_\mathrm{s}}\) is the spectral separation between two adjoining sub-carriers. Hereby, we also optimise the baseband bandwidth to improve the SE. To optimise the baseband bandwidth in HO-OFDM-IM, we employ the filtering approach proposed in \cite{mazahir2019achievable}. The details of the approach are as follows. \(x^\mathrm{c}(n)\) is first converted to \(X^\mathrm{c}[\sigma]\) using \(LN\)-order FFT which is then passed through a low-pass filter having impulse response \(H^\mathrm{F}[\sigma] = 1\) for \(\sigma \in \boldsymbol{\sigma}^+ \cup \boldsymbol{\sigma}^- \) and zero otherwise, where \(\alpha \in [ 0,(L-1)\sfrac{N}{2}]\). The low-pass filtered signal is given as \(X^\mathrm{F}[\sigma]=X^\mathrm{c}[\sigma]\times H^\mathrm{F}[\sigma]\). The bandwidth of the low-pass filter is \(B_\mathrm{F}= (N+2\alpha-1)/{T_\mathrm{s}}\), which culminates in baseband bandwidth of \(B=(\sfrac{N}{2}+\alpha)/{T_\mathrm{s}}\) for \(X^\mathrm{F}[\sigma]\). For simplicity, we shall only consider \(\alpha \in [ 0,\sfrac{N}{2}]\). It may be noticed that \(B = B_\mathrm{opt}\) when \(\alpha = 0\), however, it can be tuned between \(B_\mathrm{opt}\) and \(\tilde{B}\) by varying \(\alpha\). Using \(LN\)-order IFFT, \(X^\mathrm{F}[\sigma]\) is transformed into \(x^\mathrm{F}(n)\), for which there is a peak regrowth in the negative amplitude excursions, thus, a bias equal to \(\beta =\vert \min x^\mathrm{F}(n) \vert\) is added resulting in \(x^+(n) =x^\mathrm{F}(n)+\beta\).
\vspace{-1mm}
\section{Simulation Results and Discussions}
For comparison with HO-OFDM-IM, we use ACEO-OFDM-IM  with \(\dot{M}\)-pulse-amplitude modulation (PAM) \cite{azim2020access} and ACO-OFDM-IM with \(\tilde{M}\)-PSK \cite{bacsar2015optical} in the sequel. The choice of using ACEO-OFDM-IM is based on the fact that it is one of the best IM scheme employing the asymmetric signal with clipping for optical wireless systems, and ACO-OFDM-IM is used as benchmark because it is one classical approaches which is quite often used as reference. Furthermore, only IM schemes are employed for comparison because of aforementioned limitations of O-OFDM schemes. 
\begin{figure}[tb]\centering
\includegraphics[trim={72 0 0 0},clip,scale=1.3]{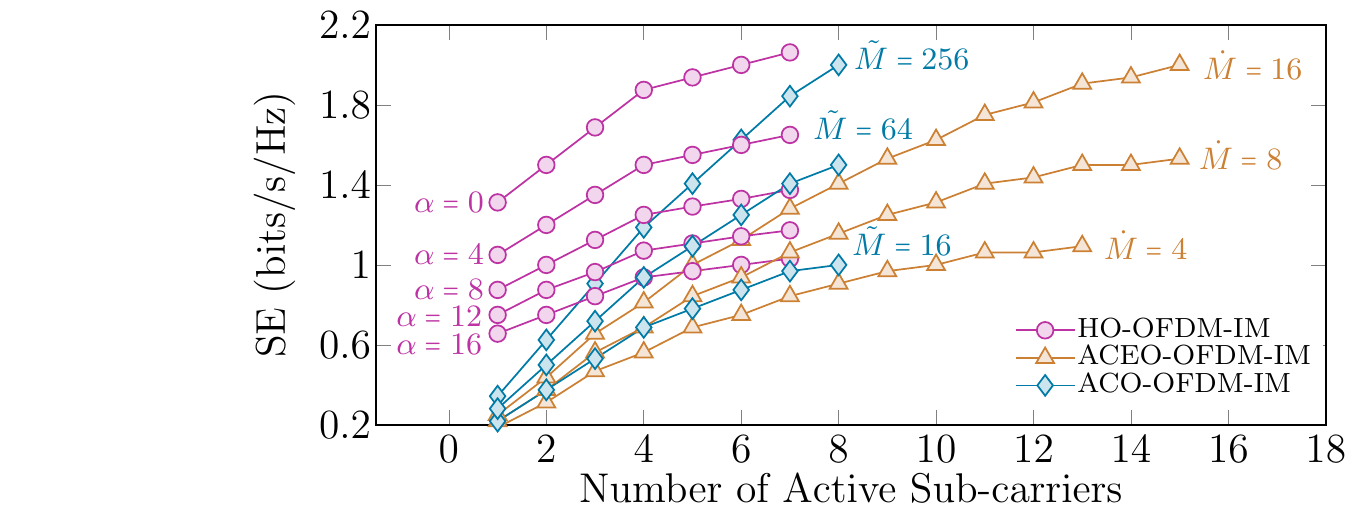}
  \caption{The evolution of SE versus the active sub-carriers for HO-OFDM-IM, ACEO-OFDM-IM and ACO-OFDM-IM, respectively for \(N=32\).}
\label{se_evolution}
\end{figure}

\begin{figure}[tb]\centering
\includegraphics[trim={18 0 7 0},clip,scale=0.7]{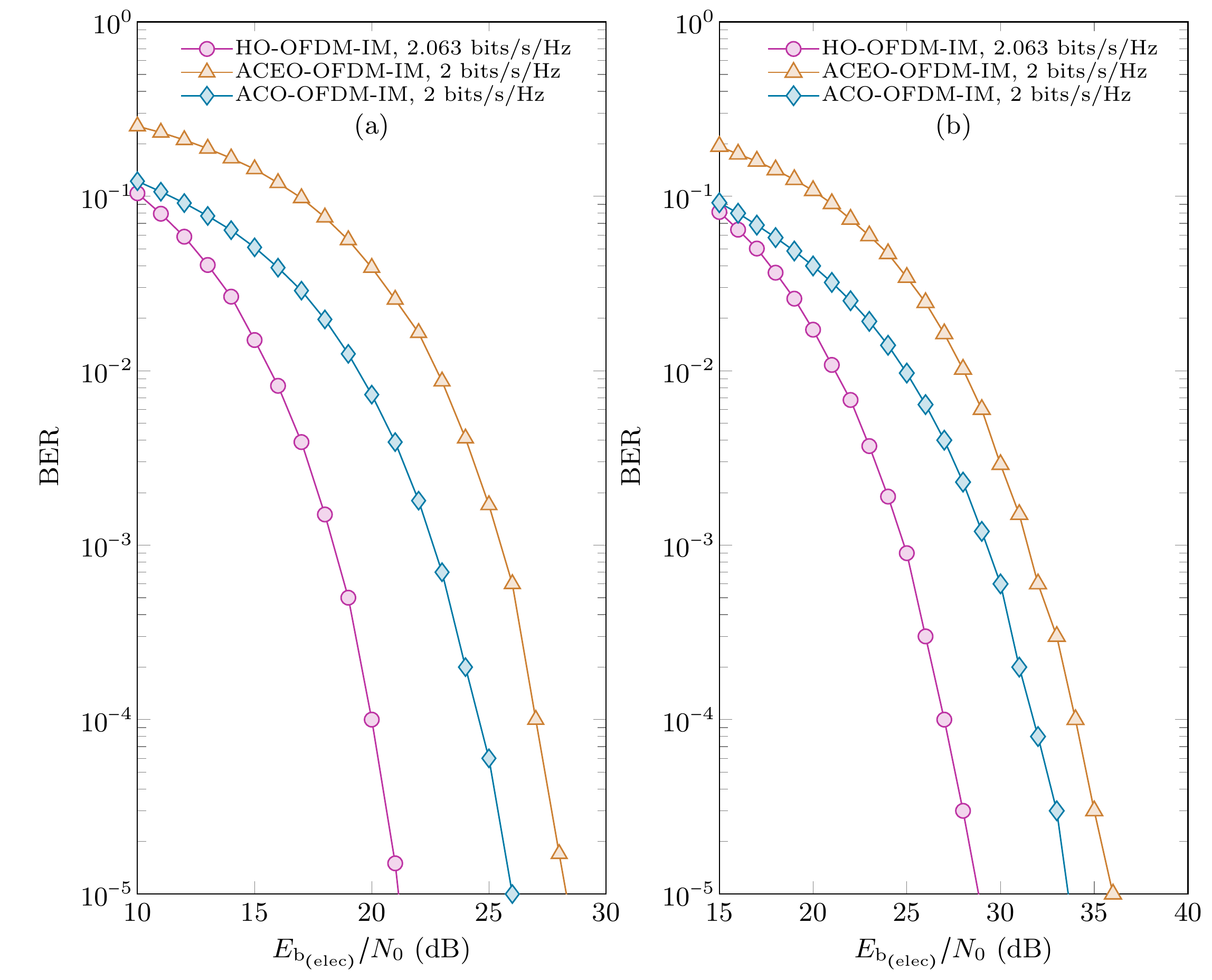}
  \caption{BER performance comparison of HO-OFDM-IM with classical counterparts for reference SE of \(2.063\) bits/s/Hz in  (a) LOS and (b) time-dispersive channel using \(N=32\). }
\label{ber}
\end{figure}

Firstly, we analyse the SE performance of HO-OFDM-IM and compare it with other benchmarks. Considering the same symbol duration, \(T_\mathrm{s}\), the evolution of spectral efficiencies of HO-OFDM-IM, ACEO-OFDM-IM and ACO-OFDM-IM for different number of active sub-carriers is presented in Fig. \ref{se_evolution}. The SE of HO-OFDM-IM is plotted considering different values of \(\alpha\). We can observe that even with QPSK, the peak SE attained by HO-OFDM-IM is around \(2.063\) bits/s/Hz, and to achieve the same SE, ACEO-OFDM-IM and ACO-OFDM-IM require \(\dot{M}=16\) and \(\tilde{M}=256\), respectively; which is not ideal in terms of EE. Another reason for better performance is that the number of active sub-carriers for HO-OFDM-IM is less compared to other schemes.
\begin{figure}[tb]\centering
\includegraphics[trim={65 0 0 0},clip,scale=1.3]{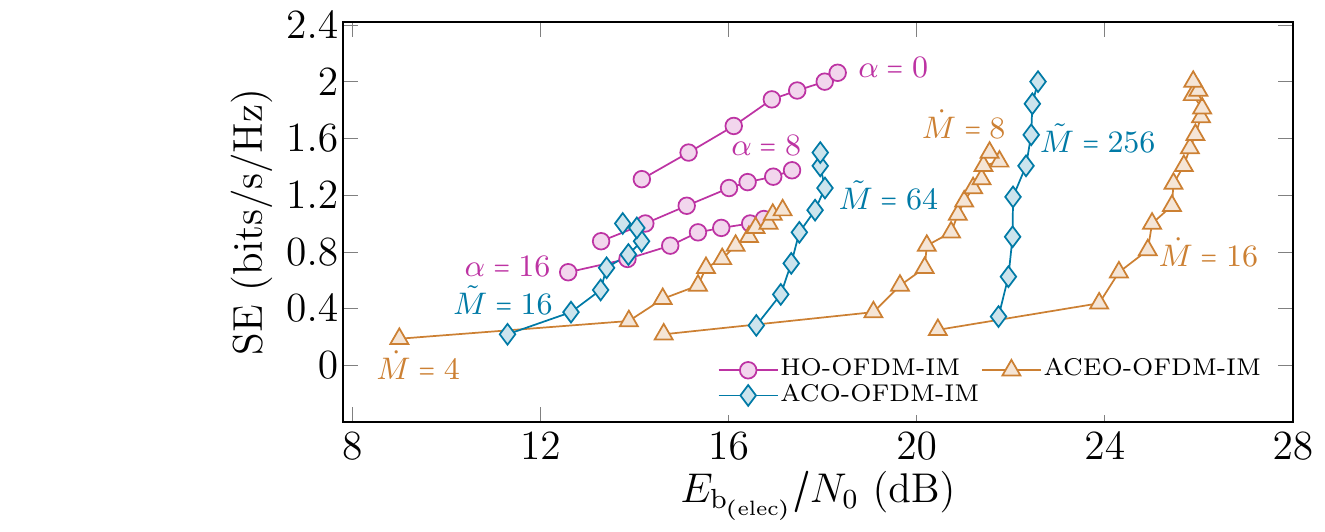}
  \caption{SE/EE trade-off for HO-OFDM-IM for different \(\alpha\) using \(N=32\) and \(L=4\) for BER of \(10^{-3}\).}
\label{se_vs_ee}
\end{figure}
Next, we move on to analysing the SE/EE trade-off offered by HO-OFDM-IM, which is illustrated in Fig. \ref{se_vs_ee}. The EE is measured by appraising the electrical signal-to-noise ratio per bit \(\sfrac{E_{\mathrm{b_{(elec)}}}}{N_0}\) needed to achieve a BER of \(10^{-3}\). \(\sfrac{E_{\mathrm{b_{(elec)}}}}{N_0}\) is equal to \(\sfrac{E_{\mathrm{s_{(elec)}}}T_\mathrm{s}}{\lambda N_0}\), where \(N_0\) is the mono-lateral noise spectral density. We adopt the energy detector based receiver for IM approaches. For HO-OFDM-IM, the SE/EE trade-off is provided for different values of \(\alpha\). It is recognised that decreasing \(\alpha\) would increase the SE increases but would also result in the reduction of EE decreases. This is because, with the reduction in \(\alpha\), the value of \(\beta\) increases following in reduction of EE. For comparison, we have chosen those alphabet cardinalities for ACEO-OFDM-IM and ACO-OFDM-IM, using which, the peak SE attained is equal to or greater than the minimum SE attained by HO-OFDM-IM with \(\alpha = 0\). HO-OFDM-IM is superior compared to other counterparts, e.g., to attain a SE of \(2\) bits/s/Hz, the \(\sfrac{E_{\mathrm{b_{(elec)}}}}{N_0}\) required by HO-OFDM-IM (with \(\alpha =0\)), ACEO-OFDM-IM and ACO-OFDM-IM is approximately \(18.04\) dB, \(25.88\) dB and \(22.58\) dB, respectively. This is because HO-OFDM-IM requires QPSK alphabets, whereas higher alphabet cardinalities are needed for ACEO-OFDM-IM, and ACO-OFDM-IM results in the reduction of EE of the respective approaches which is not optimal \cite{azim2020access}. A direct alternative of HO-OFDM-IM could be the dual-mode (DM) variants of ACEO-OFDM-IM and ACO-OFDM-IM, which are referred to as ACEO-OFDM-DM and ACO-OFDM-DM \cite{azim2020access}. However, by performing simulations under a variety of system set-ups, we determine that as they also require higher-order alphabets to attain the peak SE attained by HO-OFDM-IM, they are even less energy-efficient than ACEO-OFDM-IM and ACO-OFDM-IM. Hence, we have chosen ACEO-OFDM-IM and ACO-OFDM-IM, which provide better performance under the circumstances.
Lastly, we analyse and compare BER performance of HO-OFDM-IM in both LOS and time dispersive channels with other counterparts for the reference SE of \(2.063\) bits/s/Hz in Fig. \ref{ber}. Reference SE corresponds to the maximum achievable SE for HO-OFDM-IM with \(\alpha =0\) and \(4\)-PSK. The closest SE achievable for ACEO-OFDM-IM and ACO-OFDM-IM is \(2\) bits/s/Hz. The impulse response for time dispersive channel is attained using ceiling bounce model \cite{carruthers1997modeling}, which is given as \(h(t) = \mu\frac{6\rho^2}{(t+\rho)^7}{u(t)}\), where \(\rho = \sfrac{2H}{c}\), \(H\) is the ceiling height, \(c\) is the speed of light, \(\mu\) is the optical path loss normalised to \(1\), \(\Delta \tau = (\sfrac{\rho}{12})\sqrt{\sfrac{13}{11}}\) is the root-mean-square (rms) delay spread of the channel, which is set to \(10\) ns, and \(u(t)\) is the unit step function. We use zero-forcing equalisation and the data rate is 500 Mbits/s and it can be observed from Fig. \ref{ber} that HO-OFDM-IM significantly outperforms other counterparts for the same SE in both LOS and time-dispersive channels. This is because the bandwidth optimisation through filtering leads to the use of low order QPSK alphabets to attain high spectral efficiencies, whereas, for the alternatives, alphabets for higher cardinalities are needed.
\section{Conclusions}
In this communication, we introduced a framework for amalgamating O-OFDM and O-OFDM-IM and validated the proof-of-concept by consolidating ACO-OFDM and ACO-OFDM-IM, where the baseband bandwidth is optimised by filtering the out-of-band clipping harmonics responsible for spectral overflow. Compared to other IM alternatives, we have shown that amalgamation of O-OFDM and O-OFDM-IM can attain higher spectral efficiencies and is more energy-efficient because of the use of lower order modulation alphabets. The results show that amalgamating O-OFDM and O-OFDM-IM could be a viable alternative to other IM approaches for OWS. We anticipate that the benefits provided herein will encourage further research into the development of the optimal receiver for O-OFDM and O-OFDM-IM amalgamated symbols.
%
\bibliographystyle{unsrt}
\bibliography{biblio}
\end{document}